# Using PCA and Factor Analysis for Dimensionality Reduction of Bio-informatics Data

M. Usman Ali
Department of Computer Science
COMSATS Institute of Information Technology
Sahiwal, Pakistan

Shahzad Ahmed
Department of Computer Science
COMSATS Institute of Information Technology
Sahiwal, Pakistan

Javed Ferzund
Department of Computer Science
COMSATS Institute of Information Technology
Sahiwal, Pakistan

Atif Mehmood
Riphah Institute of Computing and Applied Sciences (RICAS)
Riphah International University
Lahore, Pakistan

Abbas Rehman
Department of Computer Science
COMSATS Institute of Information Technology
Sahiwal, Pakistan

*Abstract*—Large volume of Genomics data is produced on daily basis due to the advancement in sequencing technology. This data is of no value if it is not properly analysed. Different kinds of analytics are required to extract useful information from this raw data. Classification, Prediction, Clustering and Pattern Extraction are useful techniques of data mining. These techniques require appropriate selection of attributes of data for getting accurate results. However, Bioinformatics data is high dimensional, usually having hundreds of attributes. Such large a number of attributes affect the performance of machine learning algorithms used for classification/prediction. So, dimensionality reduction techniques are required to reduce the number of attributes that can be further used for analysis. In this paper, Principal Component Analysis and Factor Analysis are used for dimensionality reduction of Bioinformatics data. These techniques were applied on Leukaemia data set and the number of attributes was reduced from to.

## I. INTRODUCTION

Bioinformatics experiments are based on Genome, DNA, RNA and Chromosomes. Genomics plays an imperative role in this field. Huge amount of data has been produced in Genomics with a substantial portion produced in Functional Genomics (in the form of protein-protein association), Structural Genomics (in the form of 3-D structure). By using NGS (Next Generation Sequencing) technique, a lot of work has been done in the field of Microarray. This technique is helpful to identify human diseases. NGS is sequencing technique which is used to detect sequences of proteomics for next generation.

Genetic Diseases are caused by Genetic disorders that are more complex because of multiple genes interaction. These disorders are breast cancer, colon cancer, skin cancer, autism, progeria, and haemophilia. These are caused by mutation in genes or sometimes inherit from parents. Leukaemia is a cancer of blood cells that occurs due to genome abnormality. Microarray includes genes expression data that is present at large scale.

Bioinformatics data needs to be store in an efficient manner and include a lot of Attributes (Variables). The major problem is that, most of tools crash when large data stored in it.

Statistics plays superlative role in the field of Bioinformatics, Mathematics and Computer Science. It is used to extract, organise, analyse and visualise large amount of data. . For this purpose, a lot of tools like Excel, Weka, Matlab and R are available. Many Statistical tests are used for the extraction of relevant information. These are t-test, chi-squared test ($\chi2$-test), ANOVA (Analysis of Variance), Kruskal-Wallis, Friedman and PCA (Principle Component Analysis) tests [1] Statistical t-test is used to check the difference between sample Mean and hypothesised value. ANOVA is parametric (distribution) test used to check the difference of dependent variables with levels of independent variables. Kruskal-Wallis is non-parametric (distribution free) test in which assumptions are not including unlike ANOVA. Friedman test is used when there is one distributed dependent variable and one independent variable with two or many levels. It is used to check the difference in reading and math scores and writing. Chi-squared test is used to compare the observed data with data according to specific hypothesis. PCA test is used to Select/Extract relevant and specific information about variables (attributes) in large dataset. In PCA, correlation is found between principal components and original data. All of these tests applied in Statistical tool.

R is open source Statistical tool that is used for loading, extracting, interpretation and analysis of data. It includes many operations such as standard deviation, correlation, mean, variance, median, mode, graphs, plot, charts, and histograms. It automatically loads libraries and packages. It performs Machine Learning Classification and Clustering tasks very quickly and effectively.

Leukaemia data has available in enormous amount. It has four types such as CLL (Chronic Lymphocytic Leukaemia), CML (Chronic Myeloid Leukaemia), AML (Acute Myeloid Leukaemia), and ALL (Acute Lymphocytic Leukaemia). Many Homo sapiens are affected by these types.

Machine Learning plays a significant role in the Selection/Extraction and Classification of data. PCA (Principle Component Analysis) test is used for extracting relevant genes information in large Leukaemia data. Factor Analysis describes the uniqueness between many variables (attributes) in data. PCA and Factor Analysis are applied in R Statistical tool. It is powerful tool for analysis of data. Extraction of relevant genes information is very important for Machine Learning Classification.

The objectives of this article are:

- To study various features of large Bioinformatics dataset (Leukaemia)
- To apply the PCA (Principal Component Analysis) and Factor analysis statistical tests for reducing the number of attributes

The rest of the paper is organised as: Section 2 explains the related work in this field. Section 3 describe experimental setup of our work in such a way that statistical test PCA (Principal Component Analysis) and Factor analysis on large Leukaemia data in RStudio tool. Section 4 highlights on results obtained from experiment and discussion about large data analysis using statistical tool. Section 5 concludes further research work for analysis on different Bio-informatic dataset using different statistical tests.

## II. RELATED WORK

Kumar et al. [2] have developed Fuzzy kNN algorithm, providing better accuracy. They select /extract the genes with the help of t-test and classify the genes using kNN (k Nearest Neighbour) by using Leukaemia data. Leu et al. [3] have proposed analysis of genomic data with the help of sampling, in which genes are classified into three groups based on their expression level. After removing the needless groups, subsets are made by using sampling. Then kNN algorithm used to determine classification accuracy that helps to remove irrelevant subsets and $\chi 2$- test is used to find relevant genes (information) resulting in better correctness with fewer genes by using 3 bioinformatics datasets from NCBI (National Centre for Biotechnology Information). Hernandez et al. [4] have developed computational method for selection of genes. After that, they classified the genes with SVM (Support Vector Machine) Machine Learning classifier by using genetic algorithm. Leukaemia, colon cancer and lymphoma datasets are used from NCBI resulting greater accuracy. Lee et al. [5] have developed GADP (Genetic Algorithm with Dynamic Parameter setting) Algorithm that is used with the $\chi 2$-test for gene selection and SVM (support vector machine) classifier is used for effective verification of genes resulting in best accuracy with fewer genes by using 6 datasets from NCBI. Kumar et al. [6] have proposed a way in which ANOVA (Analysis of Variance) Statistical test is used for relevant gene (information) selection and kNN classifier algorithm is used for gene classification resulting in best scalability and speedup by using NCBI datasets. Ray et al. [7] have developed framework for microarray data analysis in which features/genes are selected with sf-ANOVA (single factor Analysis of Variance) and features are classified with ML (Machine Learning) techniques such Naïve Bays and Logistic Regression resulting in better scalability, correctness and speedup as compared to all existing approaches. Ali et al. [8] have explained brief description on Microarray data analysis (genes) in which many genes Selection/Extraction and Classification tests/Algorithms are discussed. They also describe the Performance comparison of different Machine Learning Techniques and Algorithms. It illustrates further research ideas in his paper about Machine Learning Techniques and Algorithms. Sarwar et al. [9] have proposed review study about Bioinformatics tools. They demonstrate the implementations of Tools for Alignment Viewers, Database Search and Genomic Analysis. It also describes further research domains for the implementation of tools using various languages such as Java, Scala, Python and R. Rehman et al. [10] have explained importance of Scala language for Bioinformatics Tools/ Algorithms. They demonstrate the supported languages for Motif Finding Tools, Multiple Sequence Alignment Tools and Pairwise Alignment tools. Ahmed et al. [11] have explains the modern data formats (models) for the implementation of Machine Learning Algorithms and techniques in Hadoop MapReduce and Spark for large Bioinformatics data. It also describes the performance comparison of different data formats. It highlights the supported platforms for different data models.

## III. EXPERIMENTS DETAIL

### A. Dataset

The Dataset used for Genome feature Selection/Extraction is obtained from NCBI (National Centre of Biotechnology Information) [12]. The details regarding this data are tabulated in TABLE I.

TABLE I. DESCRIPTION OF GENOMICS (LEUKEMIA) DATASET

| Accession | GSE13159 Family (Series Matrix File) | |
|---|---|---|
| ID_REF | From GSM329407 to GSM331732 | |
| Title | MILES stage 1 data (N1_0001 ----- N1_2096) | |
| Sample type | RNA | |
| Number of Attributes | 2096 | |
| Source name | Patient sample | |
| Total Classes | 18 | |
| Organism | Homo Sapiens (Scientific Name) | |
| Sample type | Bone Marrow | Peripheral Blood |
| WHO (World Health Organisation) Classification of Leukaemia types | Names of Classes | |
| | o mature B-ALL with t(8;14) <br> o Pro-B-ALL with t(11q23)/MLL <br> o c-ALL/Pre-B-ALL with t(9;22) <br> o T-ALL <br> o ALL with t(12;21) <br> o ALL with t(1;19) <br> o ALL with hyper-diploid karyotype <br> o c-ALL/ Pre-B-ALL without t(9;22) <br> o AML with t(8;21) <br> o AML with t(15;17) <br> o AML with inv(16)/t(16;16) <br> o AML with t(11q23)/MLL <br> o AML with normal karyotype + other abnormalities <br> o AML complex aberrant karyotype <br> o CLL <br> o CML <br> o MDS <br> o Non-Leukaemia and healthy bone marrow | |

In TABLE I four main types of Leukaemia are explained including AML, CML, ALL and CLL. In the bone marrow, Multi-potential stem cells are present. These cells are immature, undifferentiated and have no shape. They perform specific function in the human body after differentiation Stem cells are further divided into Myeloid and Lymphoid cells. Myeloid cells make Myeloblast cells which are further differentiated into Red blood cells (Erythrocyte), White blood cells and Platelets (Thrombocyte). Red blood cells are helpful to provide oxygen in human body. White blood cells defend the body from infections. Platelets provide blood clots in case of any injury [13].

AML occurs in bone marrow and blood. When immature Myeloblast cells are not differentiated then size of these cells increases. These immature Myeloblast cells get spread rapidly throughout the whole human body. Due to this reason, AML is produced. Its symptoms are fever, fatigue and bleeding. AML occurs in adults (below the age of 35 years) and children (from 2 to 9 years of age). CML occurs in bone marrow and blood. Myeloblast contains chromosomes. Genes are produced in these chromosomes. When gene for Myeloblast mutate or transfer from chromosome 9 (normal) to 22 (abnormal) then these Myeloblast cells cannot mature into RBC (Red Blood Cells), WBC (White Blood Cells) and Platelets. Due to this effect, CML is produced. Its symptoms are anaemia (due to loss of blood), fatigue and weight loss. CML occurs in elder people.

Lymphoid cells make Lymphoblast. Lymphoblast cells are further differentiated into B cells, T cells and Natural killer cells. B cells contain antibodies. When antigens enter into our body, B cells fight with them. T cells weaken the antigens and give to the B cells that remove them. If antigens are not controllable by B and T cells then Natural killer controls them.

ALL occurs in bone marrow and blood. Lymphoblast contains Lymph nodes. Lymphocyte goes to Lymph nodes to mature into B and T cells. When Lymphoblasts and Lymphocytes accumulate into Lymph nodes then ALL occurs rapidly. Its symptoms are fever, fatigue and swollen nodes (painful). CLL occurs in bone marrow and blood. When Lymphoblast has too many divisions of immature cells then CLL produces slowly. Its symptoms are anaemia and weight loss. However, CLL is less dangerous.

WHO (World Health Organisation) classification of Leukaemia types and sub-types are tabulated in TABLE I. Occurrence of 18 subclasses for Bone Marrow sample is given in Fig. 1.Similarly, occurrence of 18 subclasses for peripheral Blood sample is given in Fig. 2.

In Fig. 1, X-axis represents all subclasses for Bone Marrow and Y-axis represents total counts (existence) of Attributes (Variables) in Data GSE13159 Family. The interval between Attributes is 50 in Y-axis. Subclass "mature B-ALL with t (8; 14)" is 12-time repeats for Bone Marrow in different Attributes of original data. Total 55 Attributes represent the class "Pro-B-ALL with t (11q23)/MLL". Total 111 Attributes represent the class "c-ALL/Pre-B-ALL with t (9; 22)". Total 170 Attributes represent the class "T-ALL". Total 58 Attributes represent the class "ALL with t (12; 21)". Total 33 Attributes represent the class "ALL with t (1; 19)". Total 39 Attributes represent the class "ALL with hyper-diploid karyotype". Total 232 Attributes represent the class "c-ALL/ Pre-B-ALL without t (9; 22)". Total 35 Attributes represent the class "AML with t (8; 21)". Total 34 Attributes represent the class "AML with t (15; 17)". Total 27 Attributes represent the class "AML with inv (16)/t (16; 16)". Total 29 Attributes represent the class "AML with t (11q23)/MLL". Total 330 Attributes represent the class "AML with normal karyotype and other abnormalities". Total 46 Attributes represent the class "AML complex aberrant karyotype". No Attribute represents the class "CLL". Total 66 Attributes represent the class "CML". Total 206 Attributes represent the class "MDS". Total 73 Attributes represent the class "Non-Leukaemia and healthy bone marrow". Maximum repeated subclass "AML with normal karyotype and other abnormalities" is 330-times.

In Fig. 2, X- axis represents all subclasses for Peripheral Blood and Y-axis represents total counts (existence) of Attributes (Variables) in dataset GSE13159 Family. The interval between Attributes is 70 in Y-axis. Single Attribute represents class "mature B-ALL with t (8; 14)" in dataset. Total 15 Attributes represent the class "Pro-B-ALL with t (11q23)/MLL". Total 11 Attributes represent the class "c-ALL/Pre-B-ALL with t (9; 22)". Total 4 Attributes represent the class "T-ALL". No Attribute represents the class "ALL with t (12; 21)". Total 3 Attributes represent the class "ALL with t (1; 19)". Single Attribute represents the class "ALL with hyper-diploid karyotype". Total 5 Attributes represent the class "c-ALL/ Pre-B-ALL without t (9; 22)". Total 5 Attributes represent the class "AML with t (8; 21)". Total 3 Attributes represent the class "AML with t (15; 17)". Single Attribute represents the class "AML with inv (16)/t (16; 16)". Total 9 Attributes represent the class "AML with t (11q23)/MLL". Total 21 Attributes represent the class "AML with normal karyotype + other abnormalities". Total 2 Attributes represent the class "AML complex aberrant karyotype". Total 448 Attributes represent the class "CLL". Total 10 Attributes represent the class "CML". No Attribute represents the class "MDS". Single Attribute represents the class "Non-Leukaemia and healthy bone marrow". Maximum repeated subclass "CLL" is 448-times.

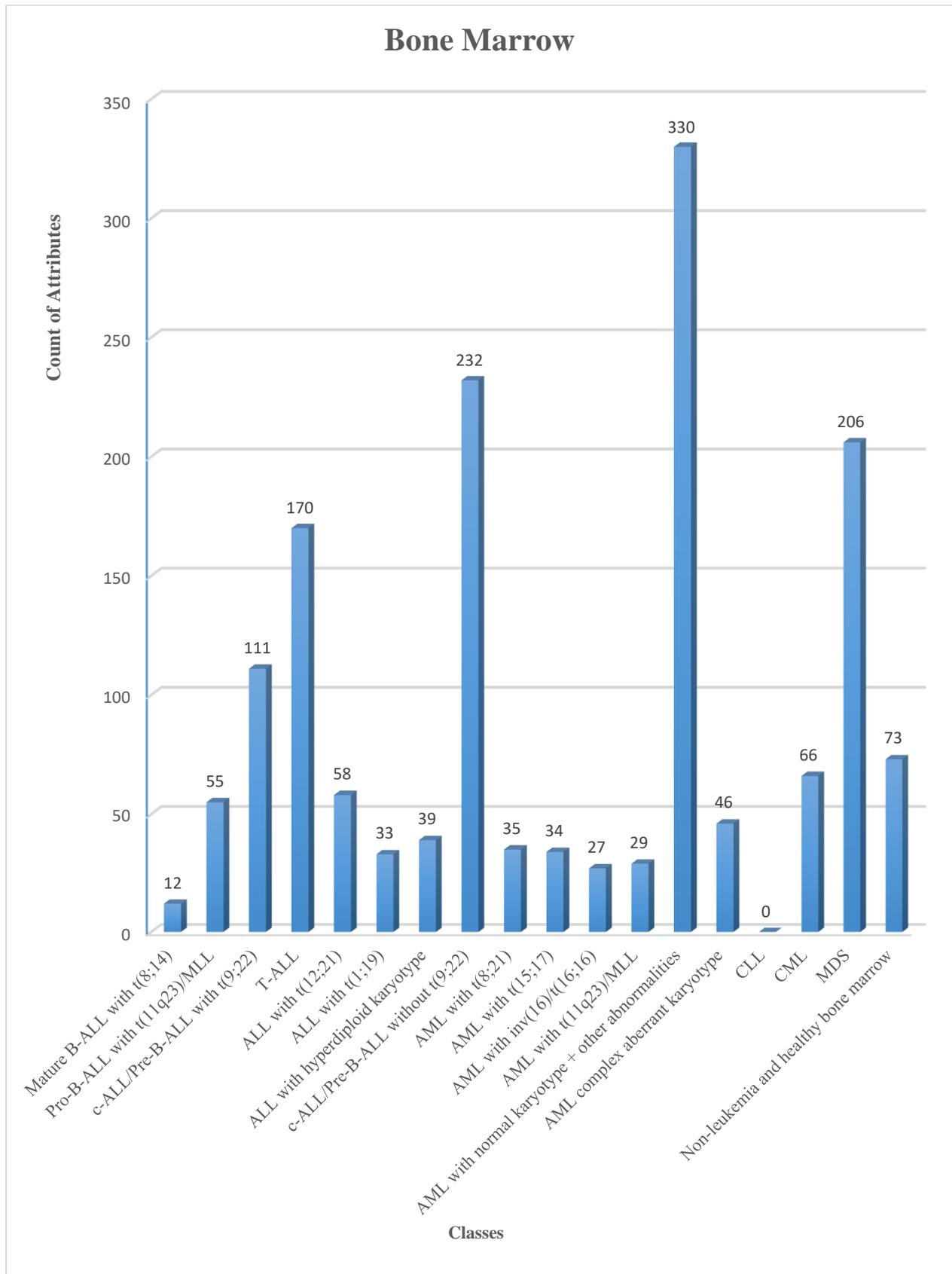

Fig. 1. Number of Attributes for subclasses in Bone Marrow sample

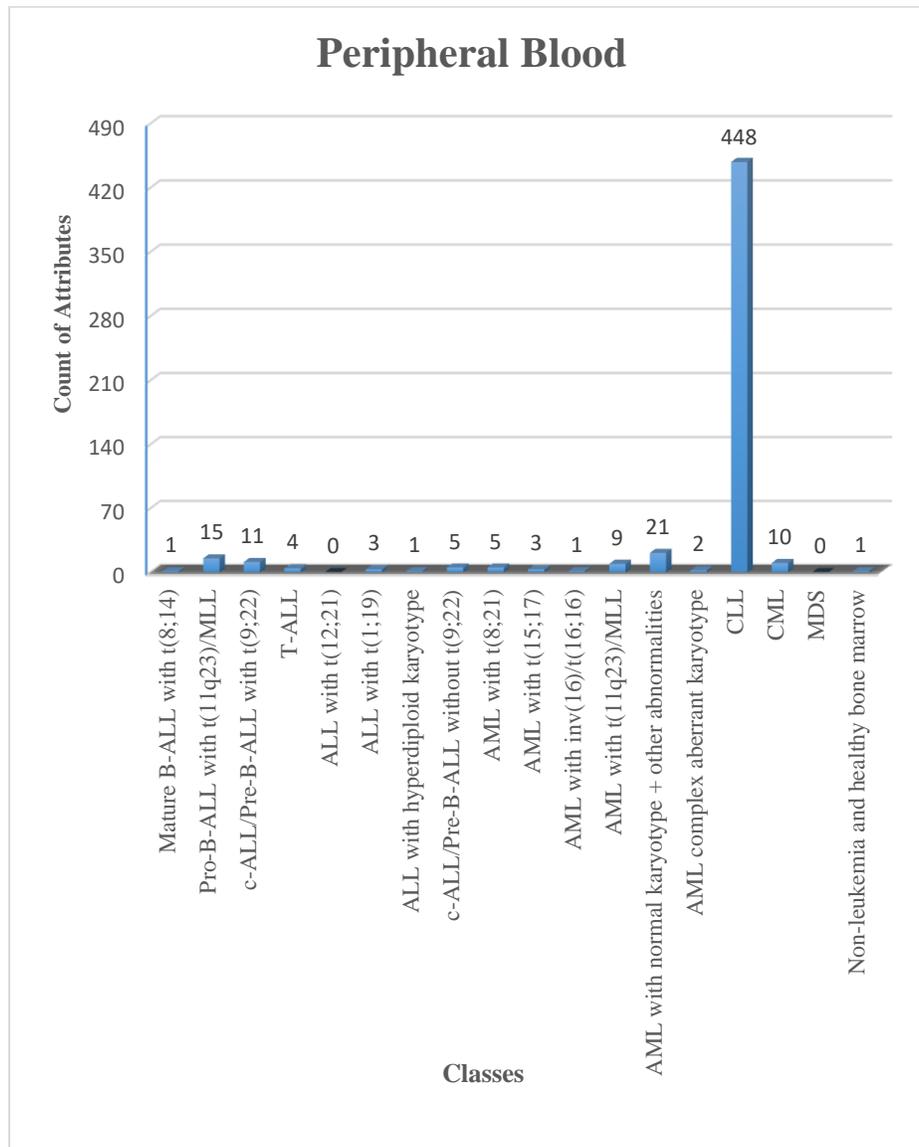

Fig. 2. Number of Attributes for sub-classes in Bone Marrow sample

### B. Preparing Dataset

Leukaemia dataset that has accession (GSE13159 Family) is too large having large number of attributes/variables. For best analysis results, whole data is divided into chunks that have equal number of attributes (variables). Every chunk has 500 numbers of attributes/values. Division of these data is shown in TABLE II.

TABLE II. DIVISION OF CHUNKS IN DATASET

| Chunks | No. of Variables | Accession no. | |
|---|---|---|---|
| | | Start | End |
| 1 | 500 | GSM329407 | GSM330130 |
| 2 | 500 | GSM330131 | GSM330636 |
| 3 | 500 | GSM330637 | GSM331136 |
| 4 | 596 | GSM331137 | GSM331732 |

### C. Statistical tool

In this experiment, 64 bit RStudio version 1.0.136 with the help of 64 bit R version 3.3.2 used [14]. PCA and Factor Analysis applied on Leukaemia dataset (GSE13159 Family (Series Matrix File)) in Statistical tool.

### D. Principal Component Analysis

Principal Component Analysis (PCA) is used to identify/ extract uncorrelated Attributes (Variables) that is called Principal Components. The main purpose of PCA is to determine maximum variance with minimum number of Principal Components [15]. In this study, PCA is applied on data that is given in TABLE I. The objective of using PCA is to reduce the dimensionality problem. PCA gives relevant gene information that is helpful for further analysis.

For Principal Component Analysis, data is loaded in R tool that reads CSV (Comma Separated Values) file. All attributes of dataset are binded in one variable by using cbind (column bind) command. Then find out correlation between all attributes. After finding correlation summary, apply PCA test. PCA scores and correlation between attributes will be true in PCA. After obtaining PCA summary, loads the final attributes. Results of PCA presented in plots, sceeplots and also in biplots.

*E. Factor Analysis*

Factor Analysis is used to find out the uniqueness among many attributes (variables). A lot of attributes exist in large Dataset. Some attributes/records are meaningless for the purpose of analysis. So observed attributes (variables) are selected with many traditional analysis techniques but these techniques do not perform well at some extent. To remove this bottleneck, Factor Analysis approach is used that finds meaningful observed attributes (variables) in large. This approach is superlative for large data analysis.

In this experiment, perform statistical PCA (Principal Component Analysis) test for extracting relevant information for dataset in R tool. Then apply Factor analysis for the uniqueness of observed attributes (variables) and extract relevant features.

For Factor Analysis, data is loaded in R tool that reads CSV (Comma Separated Values) file. So, we need frames of standardised attributes/variables for further processing. For this, convert whole data into specific frame. Then apply Factor analysis using command factanal ( ). By using this command, find out the results initially without rotation of attributes/records. After Factor analysis, find 10 factors. Important arguments of Factor analysis are dataset, number of factors, rotation that will be none initially and omits null values of attributes/variables. After loading the results of Factor Analysis, compute Eigen values. Then find out the proportion of variance of Eigen values. Now, compute the uniqueness among different attributes/variables. Finally, resultant graph generated without factor rotation. Next, find out Factor analysis using varimax rotation. After loading factor variables, draw resultant graph with varimax factor rotation in [Fig. 4] – [Fig. 7].Then find out the variables that has minimum and maximum values of factor 1 and factor 2 [16]. After binding these selected variables of factor 1 and factor 2, generates resultant graph with selected attributes/variables in [Fig. 8] – [Fig. 11].

IV. RESULTS AND DISCUSSION

Bioinformatics field consists of proteins, genes, DNA, RNA and chromosomes. It also contains data of Leukaemia disease which occurs in multiple forms such as CLL, CML, AML and ALL. All of these types occur due to large number of genes in human body. These data need to be analyses in an effective and efficient manner. A lot of statistical tools are used for analysis of these data but PCA (Principal Component Analysis) test and Factor analysis are more preferable.

In this experiment, large Leukaemia data is used that is divided into chunks and analysed every chunk. Statistical PCA (Principal Component Analysis) test is applied on given data. PCA test applied on every chunk that has the same number of attributes/variables. In Fig. 3 after performing PCA, when load attributes of whole dataset, it gives only 9 components among 500 components. The reason is that these resultant 9 components have greater than one value. The remaining components have less than one value. Results for analysis are represented using graph of PCA test.

Factor analysis test applied on every chunk that has the same number of attributes/variables. After loading data in R, compute Eigen values and communality distance among variables/attributes of given dataset. Then check the uniqueness of the variables and perform Factor analysis without rotation of factors. Representation of attributes/variables that have minimum and maximum values for factor 1 and factor 2 is given in [Fig. 4] – [Fig. 7].

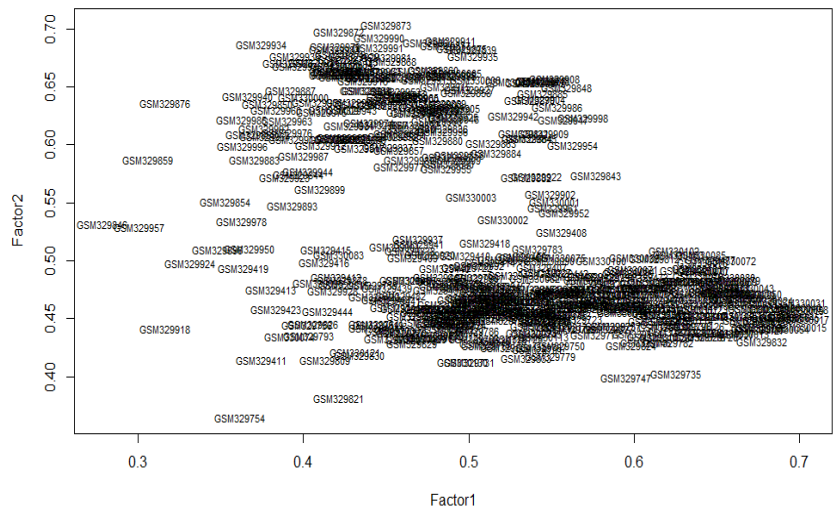

Fig. 3. Plot of Attributes for Factor Analysis with Rotation of Factors (1-500)

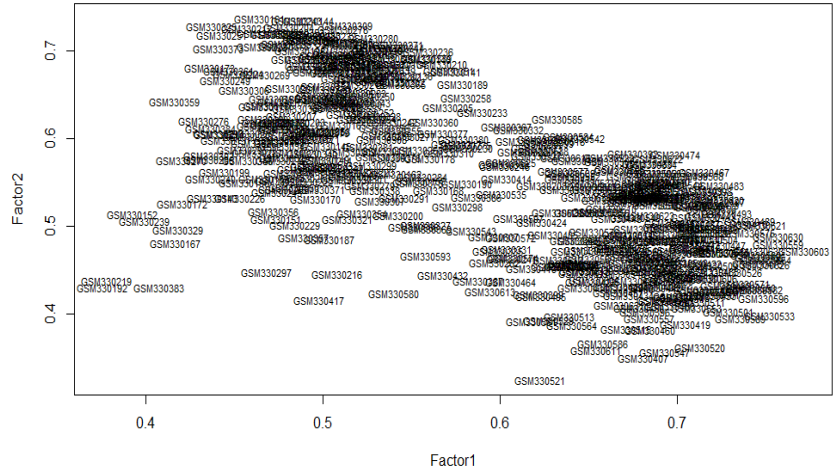

Fig. 4. Plot of Attributes for Factor Analysis with Rotation of Factors (500-1000)

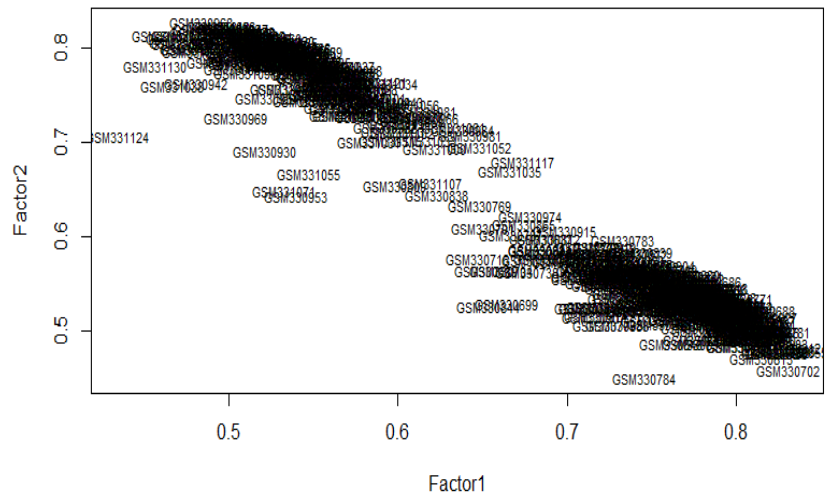

Fig. 5. Plot of Attributes for Factor Analysis with Rotation of Factors (1000-1500)

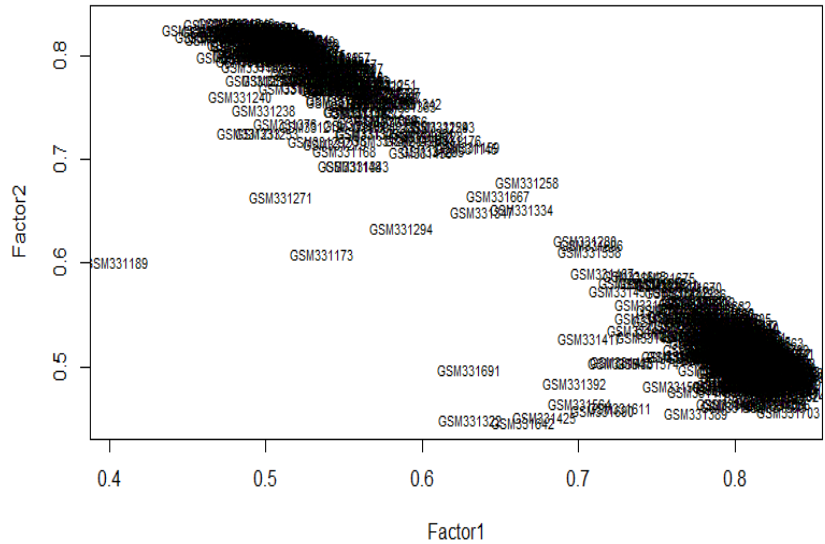

Fig. 6. Plot of Attributes for Factor Analysis with Rotation of Factors (1500-2096)

When extracted variables/attributes with rotation of factor are gained, then bind these variables with randomly selected other variables in given specific dataset. Finalised extracted attributes/variables using Factor analysis are shown in [Fig. 8] – [Fig. 11].

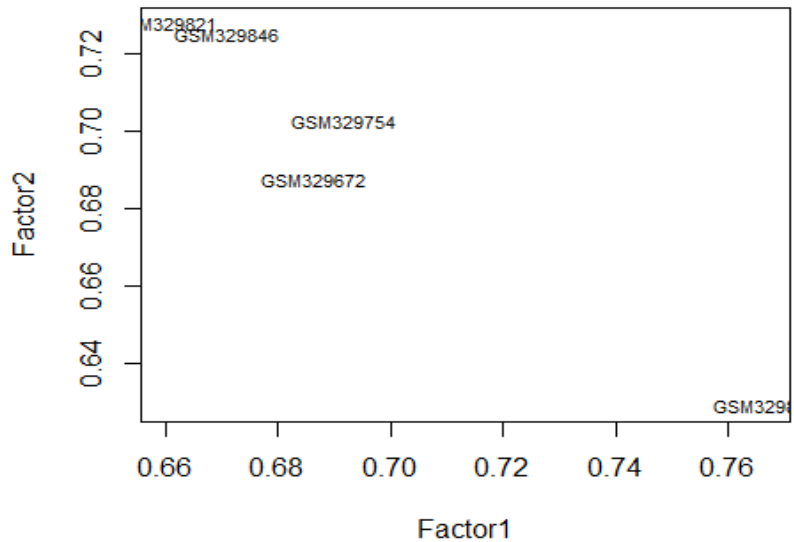

Fig. 7. Finalised Extracted Attributes using Factor Analysis (1-500)

In Fig. 8, the finalised Attributes are GSM329821, GSM329846, GSM329754, GSM329672 and GSM329946. GSM329821, GSM329846, GSM329754 and GSM329946 Attributes/Variables represent Bone Marrow sample but subclass varies. GSM329821 and GSM329754 Attributes have c-ALL/Pre-B-ALL with t (9; 22) subclass. GSM329846 Attribute has T-ALL subclass.

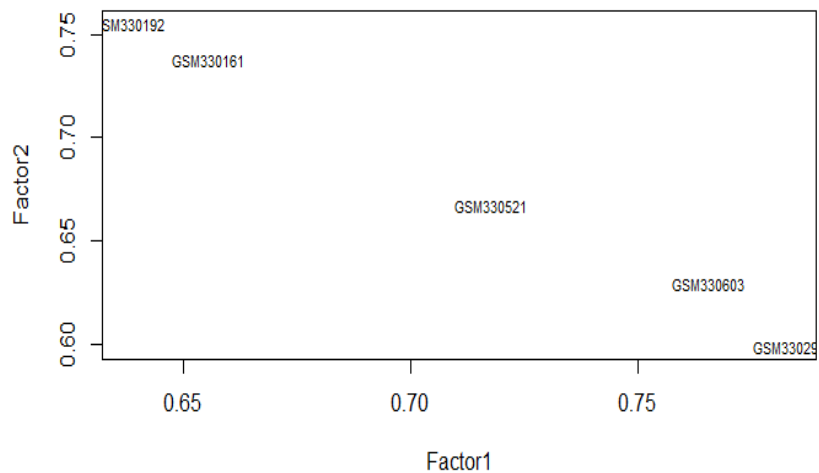

Fig. 8. Finalised Extracted Attributes using Factor Analysis (500-1000)

In Fig. 9. the finalised Attributes are GSM330192, GSM330161, GSM330521, GSM330603 and GSM330291. GSM330192, GSM330161, GSM330291 and GSM330603 Attributes/Variables represent Bone Marrow sample in but GSM330521 represent Peripheral Blood sample. GSM330192, GSM330291 and GSM330161 attributes have c-ALL/Pre-B-ALL without t (9; 22) subclass. GSM330521 Attribute has AML with t (11q23)/MLL subclass. GSM330603 Attribute has AML with normal karyotype and other abnormalities subclass.

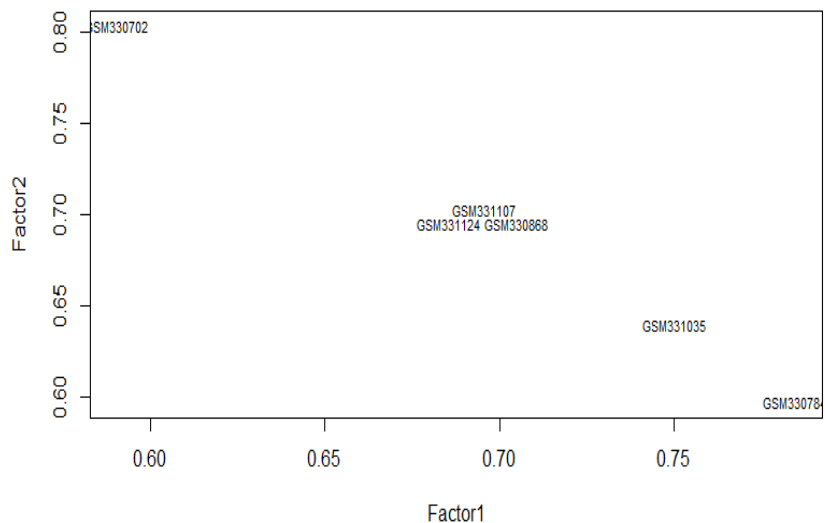

Fig. 9. Finalised Extracted Attributes using Factor Analysis (1000-1500)

In Fi.g. 10, the finalised Attributes are GSM330702, GSM331107, GSM331124, GSM330868, GSM331035 and GSM330784. GSM331107, GSM331124 and GSM331035 Attributes/Variables represent Peripheral Blood sample but GSM330702, GSM330784 and GSM330868 represent Bone Marrow sample. GSM331107, GSM331124 and GSM331035 attributes have CLL subclass. GSM330702, GSM330868 and GSM330784 Attributes have AML with normal karyotype and other abnormalities subclass.

In Fig. 11, finalised Attributes are GSM331189, GSM331703, GSM331675, GSM331258, GSM331240, GSM331657, GSM331173, GSM331692 and GSM331547. GSM331189, GSM331258, GSM331240 and GSM331173 Attributes/Variables represent Peripheral Blood sample.

GSM331703, GSM331675, GSM331657, GSM331547 and GSM331692 represent Bone Marrow sample. GSM331189, GSM331258, GSM331240 and GSM331173 attributes have CLL subclass. GSM331703, GSM331675 and GSM331692 Attribute have Non-leukaemia and healthy bone marrow subclass. GSM331173 and GSM331547 Attributes have MDS subclass.

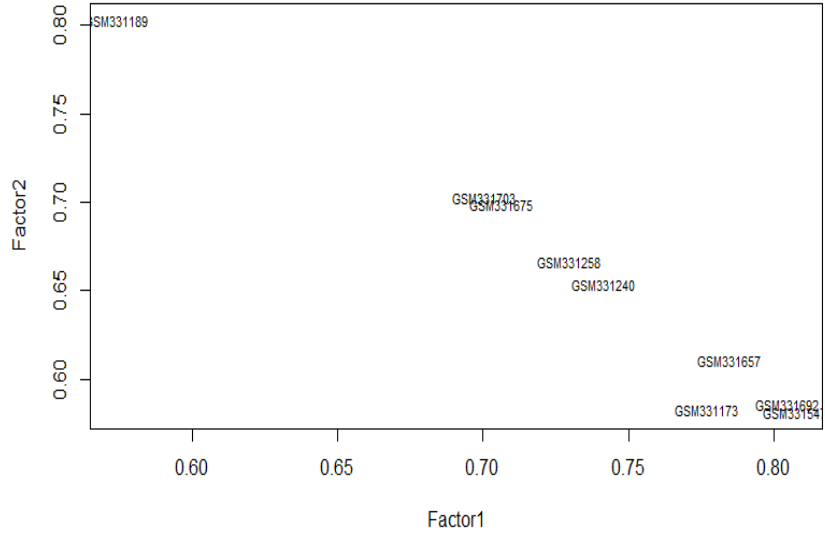

Fig. 10. Finalised Extracted Attributes using Factor Analysis (1500-2096)

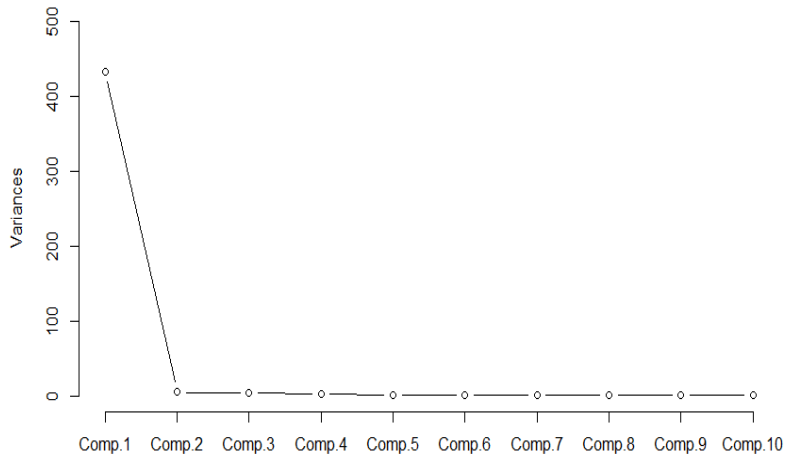

Fig. 11. Scree Plot of Attributes for PCA

## V. CONCLUSION

Accuracy of data mining experiments depends upon the appropriate selection of attributes for analysis. Further, larger the number of attributes, more time and space will be required for processing the data. Bio-informatics data usually have high dimensions which need to be reduced for applying machine learning algorithms. Statistical techniques are available for dimensionality reduction and selection of features. In this paper, a study was presented on reducing the number of attributes using PCA and Factor Analysis. Leukaemia data set was used for the experiments. First, PCA was applied on the data set and 9 components were selected out of the 500 components. Then Factor Analysis was used to extract the important features. GSM330702, GSM331107, GSM331124, GSM330868, GSM331035 and GSM330784 are found to be the important attributes in Leukaemia data.

In future, results of this study will be used for classification and prediction experiments.

REFERENCES

[1] "What statistical analysis should I use?Statistical analyses using SAS - IDRE Stats," [Online]. Available: http://stats.idre.ucla.edu/sas/whatstat/what-statistical-analysis-should-i-usestatistical-analyses-using-sas/.

[2] Mukesh Kumar, Santanu Ku. Rath, "Microarray Data Classification using Fuzzy K-Nearest Neighbor," 2014.

[3] Yungho Leu, Chien-Pang Lee, and Hui-Yi Tsai, "A Gene Selection Method for Microarray Data Based on Sampling," in ICCCI, Verlag Berlin Heidelberg, 2010.

[4] Jose Crispin Hernandez Hernandez , Jin-Kao Hao, B´eatrice Duval;, "A Genetic Embedded Approach for Gene Selection and Classification of Microarray Data," in EvoBio, Verlag Berlin Heidelberg, 2007.

[5] Chien-Pang Lee , Yungho Leu, "A novel hybrid feature selection method for microarray data analysis," Applied Soft Computing, vol. 11, no. 1, pp. 208-213, January 2011.

[6] N. K. R. A. S. S. K. R. Mukesh Kumar, "Feature Selection and Classification of Microarray Data using MapReduce based ANOVA and K-Nearest Neighbor," in IMCIP, 2015.

[7] M. K. S. K. R. Ransingh Biswajit Ray, "Fast Computing of Microarray Data Using Resilient Distributed Dataset of Apache Spark," in Recent Advances in Information and Communication Technology, vol. 463, Springer International Publishing, 2016, pp. 171-182.

[8] M. U. Ali, S. Ahmad and J. Ferzund, "Harnessing the Potential of Machine Learning for Bioinformatics using Big Data Tools," International Journal of Computer Science and Information Security (IJCSIS), vol. 14, no. 10, pp. 668-675, 2016.

[9] M. A. Sarwar, A. Rehman and J. Ferzund, "Database Search, Alignment Viewer and Genomics Analysis Tools: Big Data for Bioinformatics," International Journal of Computer Science and Information Security (IJCSIS), vol. 14, no. 12, pp. 317-328, 2016.

a. Rehman, A. Abbas, M. A. Sarwar and J. Ferzund, "Need and Role of Scala Implementations in Bioinformatics," International Journal of Advanced Computer Science and Applications (IJACSA), vol. 08, no. 02, 2017.

[10] S. Ahmed, M. U. Ali, J. Ferzund, M. A. Sarwar, A. Rehman and A. Mehmood, "Modern Data Formats for Big Bioinformatics Data Analytics," International Journal of Advanced Computer Science and Applications (IJACSA), vol. 8, no. 4, 2017.

[11] "Download data for GSE13159 - GEO - NCBI," [Online]. Available: https://www.ncbi.nlm.nih.gov/geo/download/?acc=GSE13159.

[12] "Types of Leukemia: 4 Primary Types | CTCA," [Online]. Available: http://www.cancercenter.com/leukemia/types/.

[13] "Download R-3.3.2 for Windows. The R-project for statistical computing.," [Online]. Available: https://cran.r-project.org/bin/windows/base/.

[14] "What is principal components analysis? - Minitab," [Online]. Available: http://support.minitab.com/en-us/minitab/17/topic-library/modeling-statistics/multivariate/principal-components-and-factor-analysis/what-is-pca/.

[15] "Factor Analysis," [Online]. Available: http://web.stanford.edu/class/psych253/tutorials/FactorAnalysis.html.